\documentclass[3p,times,twocolumn]{elsarticle}
\biboptions{comma,sort&compress}

\usepackage{color} 
\usepackage[normalem]{ulem} 

\usepackage{graphicx}
\usepackage{amsmath}
\usepackage{ecrc}
\usepackage[rflt]{floatflt}
\usepackage{float}
\allowdisplaybreaks[4]
\newcommand{\Fh}[2]{\,{}_#1F_#2}
\newcommand{\Fs}[3]{\!\!\left[\begin{array}{c}#1\,;\\#2\,;\end{array}#3\right]}
\newcommand{\Fz}[3]{\Fs{#1}{#2}{#3}}

\volume{00}
\firstpage{1}
\journalname{Nuclear and Particle Physics Proceedings}
\runauth{}
\jid{nppp}
\jnltitlelogo{Nuclear and Particle Physics Proceedings}
\usepackage{amssymb}
\usepackage[figuresright]{rotating}
\begin{document}
\begin{frontmatter}
\title{{\footnotesize DESY 15-159, DO-TH 15/09, {\tt arXiv:1510.01063[hep-ph]}} \\
General $\varepsilon$-representation for scalar one-loop Feynman integrals}
\author{Johannes Bl\"umlein}
   \ead{johannes.bluemlein@desy.de}
\author{Khiem Hong Phan\fnref{fn1}}
\fntext[fn1]{Speaker, corresponding author.}
   \ead{khiem.hong.phan@desy.de}
\author{Tord Riemann}
   \ead{tord.riemann@desy.de}
\address{Deutsches Elektronen-Synchrotron DESY, Platanenallee 6, 15738 Zeuthen, Germany}
\pagestyle{myheadings}
\markright{ }
\begin{abstract}
\noindent
A systematic study of the scalar one-loop two-, three-, and four-point Feynman integrals is performed.  
We consider all cases of mass assignment and external invariants and derive closed expressions in arbitrary 
space-time dimension in terms  of higher transcendental functions. The integrals play a role as building 
blocks in general higher-loop or multi-leg processes. We also perform numerical checks of the calculations 
using {\tt AMBRE/MB} and {\tt LoopTools/FF}.
\end{abstract}
\begin{keyword} 
One-loop Feynman integrals, (generalized) hypergeometric functions.
\end{keyword}
\end{frontmatter}
\section{Introduction}

\noindent
The physics at future colliders, like the LHC at high luminosities and the 
ILC~\cite{ATLAS:2013hta,CMS:2013xfa,Baer:2013cma}, focuses on measuring 
the properties of the Higgs boson, of the top quark and vector bosons, as well as
performing searches for signals beyond the Standard Model. These measurements
will be performed at high precision, requiring higher order corrections.
They necessitate detailed 
calculations for one-loop multi-leg processes and higher-loop calculations for selected scattering cross sections.

Scalar one-loop integrals in general space-time dimension $d = 4+ 2n -2 \varepsilon,
~n \in \mathbb{N}$, are important for several reasons. Within the general framework 
of higher-order corrections, higher terms in the  $\varepsilon$-expansion for one-loop 
integrals form necessary building blocks.
At the level of one-loop order, one may cure the inverse Gram determinant problem 
with Feynman integrals in higher dimensions than 
$d=4$~\cite{Davydychev:1991va,Bern:1993kr,Campbell:1996zw,Binoth:2005ff,Fleischer:2010sq}.  

The calculation of scalar one-loop integrals in $d = 4 - 2 \varepsilon$ dimensions 
have been performed by many authors~\cite{'tHooft:1978xw,vanOldenborgh:1989wn, 
Denner:1991qq, Nhung:2009pm, Denner:2010tr,Bauer:2001ig, Do:2010zz}. For the case of general
dimension $d$, we mention the work in ~\cite{Boos:1990rg, Davydychev:1990cq, 
Davydychev:1997wa, Tarasov:2000sf, Fleischer:2003rm}. Various packages are available
for the numerical evaluation of massive one-loop integrals, such as {\tt FF}~\cite{vanOldenborgh:1990yc}, 
{\tt LoopTools}~\cite{Hahn:1998yk}, {\tt XLOOP-GiNaC}~\cite{Bauer:2001ig},  
{\tt AMBRE/MB}~\cite{ambre
,Czakon:2005rk}
 and others \cite{OTH}. However, not all of these calculations and 
packages 
cover general dimension $d$ with a 
general $\varepsilon$-expansion. In  this paper,
we study systematically the scalar one-loop 2-, 3-, and
4-point integrals, based on the method introduced in~\cite{ Tarasov:2000sf, Fleischer:2003rm}. We consider all 
cases of mass 
and external invariant assignment with propagator indices $\nu_k = 1$ and perform numerical checks using 
{\tt LoopTools/FF} and {\tt AMBRE/MB}. 
\section{The one-loop functions}

\noindent
In the following we present analytic results for scalar one-loop two-, three-, and four-point functions  at general
values of space-time dimension $d$.
\subsection{Notations}%

\noindent
Scalar one-loop $n$-point Feynman integrals are given by
\begin{eqnarray}
\label{npoint}
J_{n}^{(d)} = \int \dfrac{d^d k}{i \pi^{d/2}} \dfrac{1}{P_1 P_2\dots P_n},
\end{eqnarray}
where $P_j = (k-p_j)^2-m_j^2$, $p_{ij}^2=(p_i-p_j)^2$  will denote the 
external momenta and $m_j$ is the mass of the $j$th propagator.

The following recurrence relation for  $J_{n}^{(d)}$ in general space-time 
dimension 
\begin{eqnarray}
\label{recurI0}
\hspace{-0.5cm} (d-n+1)\; G_{n-1} J^{(d+2)}_n = \Big[2 \Delta_n 
     + \sum\limits_{k=1}^{n}(\partial_k \Delta_n) \; \mathbf{k}^-\Big]J^{(d)}_n
\end{eqnarray}
has been given~\cite{Fleischer:2003rm}. Here $d$ is the space-time dimension, $\partial_k \equiv 
\partial / \partial m_k^2$, and $\mathbf{k}^- $ is an operator which 
shrinks $P_k$ in the integrand of $J_n^{(d)}$.  The kinematic 
variables in ~(\ref{recurI0}) are defined as follows 
\begin{eqnarray}
\hspace{-0.5cm}
\label{yijk}
\Delta_n(\{p_1,m_1\},\ldots \{p_n,m_n\}) =  \left|
\begin{array}{cccc}
Y_{11}  & Y_{12}  &\ldots & Y_{1n} \\
Y_{12}  & Y_{22}  &\ldots & Y_{2n} \\
\vdots  & \vdots  &\ddots & \vdots \\
Y_{1n}  & Y_{2n}  &\ldots & Y_{nn}
\end{array}
         \right|,~~~\nonumber
\label{deltan}
\end{eqnarray}
\begin{eqnarray}
\hspace*{-0.7cm}
\text{with}~~
Y_{ij}=-(p_i-p_j)^2+m_i^2+m_j^2,
\end{eqnarray}
and
\begin{eqnarray}
\label{Gram}
&&\hspace{-1.2cm} G_{n-1}(p_1,\ldots ,p_n) = \nonumber \\
&&\hspace{-0.5cm}= -2^n \left|
\begin{array}{cccc}
  \! p_1^2 & p_1p_2   &\ldots & p_1p_{n-1} \\
  \! p_1p_2  &  p_2^2 &\ldots & p_2p_{n-1} \\
  \vdots  & \vdots  &\ddots   & \vdots \\
  \!p_1p_{n-1}  & p_2p_{n-1}  &\ldots & p_{n-1}^2
\end{array}
\right|. ~~
\end{eqnarray}

As will be discussed later, the analytic results for scalar one-loop
$n$-point integrals will be presented as a function of the ratio 
of the above determinants. Therefore, it is worth to introduce the 
following index variables
\begin{eqnarray}
\lambda_{ij\cdots n}  &=&  \Delta_n(\{p_1,m_1\},\ldots \{p_n,m_n\}), 
                                                         \\
       g_{ij\cdots n} &=& G_{n-1}(p_1,\ldots ,p_n),
\\
r_{ij\ldots n}&=&-\frac{\lambda_{ij\ldots n}}{g_{ij\ldots n}}.
\end{eqnarray}

The solution of ~(\ref{recurI0}) has been presented in~\cite{Fleischer:2003rm}:
\begin{eqnarray}
\label{npoint}
&&\hspace{-1cm} J_n^{(d)}=b_n(\varepsilon)-                              \\
&&\hspace{-0.9cm} -\sum_{k=1}^{n} \left(\frac{\partial_k \Delta_n}
   {2 \Delta_n}\right)\sum_{r=0}^{\infty} \left(\frac{d-n+1}{2}
   \right)_r\left(\frac{G_{n-1}}{\Delta_n}\right)^{r}
   {\bf k^{-} } J^{(d+2r)}_n.                                   \nonumber
\end{eqnarray}
The boundary term $b_n(\varepsilon)$ is determined by the asymptotic 
behavior $J_n^{(d)}$ at $d \rightarrow \infty$.  
\subsection{Two-point functions}

\noindent
We first consider the simplest case, the scalar one-loop two-point functions
for $p_{ij}^2 \neq 0$ and $r_{ij} \neq 0$. The recurrence relation for 
$J_2^{(d)}$ takes the form
\begin{eqnarray}
&&\hspace{-1cm}J_2^{(d)}  = b_2(\varepsilon)-\nonumber\\
&&\hspace{-0.5cm}-\sum_{k=1}^{2} 
\left(\frac{\partial_k \lambda_{ij}} {2 \lambda_{ij}}\right)
\sum_{r=0}^{\infty} \left(\frac{d-1}{2} \right)_r
\left(-\frac{1}{r_{ij}}\right)^{r}{\bf k^{-} } J^{(d+2r)}_2
\label{2point}
\end{eqnarray}
and $k =1,2$ label the internal masses. 
The operator ${\bf k^{-} } J^{(d+2r)}_2$  reduces $J_2^{(d)}$ 
to scalar one-loop one-point functions in  $d+2r$ space-time 
dimensions. By using the formula for the scalar one-loop one-point 
functions in \cite{Fleischer:2003rm} one obtains
\begin{eqnarray}
{\bf k^{-} } J^{(d+2r)}_2  = 
  (-1)^{r+1} \dfrac{\Gamma(\frac{d}{2}) \Gamma(1-\frac{d}{2})}{\Gamma(\frac{d}{2}+r)}
  (m_k^2 )^{\frac{d-2}{2} +r}.
\end{eqnarray}
Inserting this result into ~(\ref{2point}), the following representation is obtained
\begin{eqnarray}
\frac{2 \lambda_{ij} J_2^{(d)} }{\Gamma\left(1-\frac{d}{2}\right)} =b_2(\varepsilon)
&+& \frac{\partial_i \lambda_{ij}}{(m_j^2)^{1-\frac{d}{2}}}
    \sum_{r=0}^{\infty} \frac{\left( \frac{d-1}{2} \right)_r}{\left(\frac{d}{2}\right)_r}
    \left(\frac{m_j^2}{r_{ij}}\right)^r               \nonumber\\ 
&&\hspace{-1.8cm}+\frac{\partial_j \lambda_{ij}}{(m_i^2)^{1-\frac{d}{2}}}
    \sum_{r=0}^{\infty} \frac{ \left(\frac{d-1}{2}\right)_r}
    {\left(\frac{d}{2}\right)_r}\left(\frac{m_i^2 }{r_{ij}} \right)^r .
\label{i2sums}
\end{eqnarray}
Here $b_2 (\varepsilon)$ is  determined by the asymptotic 
behavior of $J_2^{(d)}$,~$d\rightarrow \infty$.
The infinite series is given in terms of a Gauss hypergeometric function ~\cite{Slater}, yielding
\begin{eqnarray}
\label{regionI}
&&\hspace{-1.3cm} 
\frac{2 \lambda_{ij}~ J_2^{(d)} }{\Gamma\left(1-\frac{d}{2}\right)} =
     -\frac{\sqrt{\pi}\Gamma\left(\frac{d}{2}\right)} {\Gamma\left(\frac{d-1}{2}\right)}\;
        r_{ij}^{\frac{d-2}{2}} \Bigg[ \frac{\partial_i  \lambda_{ij}}
        {\sqrt{1-\frac{m_j^2}{r_{ij}}}}
       +\frac{\partial_j \lambda_{ij}}
        {\sqrt{1-\frac{m_i^2}{r_{ij}}}} \Bigg] \nonumber \\
&& \nonumber \\
&&\hspace{-1.2cm}
       +~\Biggl\{  \frac{\partial_i \lambda_{ij}}{(m_j^2)^{1-\frac{d}{2}}}
         \Fh21 \; \Fz{1,\frac{d-1}{2}}{\frac{d}{2}} {\frac{m_j^2}{r_{ij} }}
       + (i \leftrightarrow j) {\Biggr\}}, 
\end{eqnarray}
provided that Re$\Big((d-1)/{2}\Big)>0$ {and} 
$\Big| {m_{i,j}^2}/{r_{ij} } \Big|<1$. 
If one applies Eqs.~(1.3.13),~ (1.8.10),~ (1.3.3.5) 
of~\cite{Slater}, one obtains
\begin{eqnarray}
\label{regionIII}
&&\hspace{-1.3cm} \frac{g_{ij}~ J_2^{(d)} }{\Gamma\left(2-\frac{d}{2}\right)} =
-  \frac{\partial_i \lambda_{ij}}{ (m_j^2)^{\frac{4-d}{2} } }
  \Fh21 \; \Fz{1, \frac{4-d}{2}}{\frac{3}{2} } {1-\frac{r_{ij}}{ m_j^2} } \\
&&\hspace{1cm}- \frac{\partial_j \lambda_{ij}} {(m_i^2)^{\frac{4-d}{2}} } 
  \Fh21 \; \Fz{1, \frac{4-d}{2}}{\frac{3}{2} }{1-\frac{r_{ij}}{m_i^2} },\nonumber
\end{eqnarray}
provided that {$\Big|1-\frac{r_{ij}}{m_{i,j}^2} \Big|<1$}.
Eqs.~(\ref{regionI}) and ~(\ref{regionIII}) reproduce (53) and  (59) in ~\cite{Fleischer:2003rm}. 

It is important to note that the arguments of the hypergeometric 
functions in ~(\ref{regionI}) and ~(\ref{regionIII})  
may have a different behavior. In general, it is not possible to write a single
expression for {$J_{n}^{(d)}$}; one rather has to refer to the corresponding analytic continuations, cf.
e.g. ~\cite{Slater}. We {will} treat all the special cases such as
$r_{ij} = \{0, m_i^2,m_j^2\}$, $g_{ij} = 0$, 
$m_i^2=m_j^2=0$, etc., in ~\cite{higheps}. In {the present} paper, 
we consider $r_{ij} =0$ as an example. One has
\begin{eqnarray}
\hspace{-0.5cm}\Fh21 \;\; \Fz{1, \frac{4-d}{2}}{\frac{3}{2} }{1 } 
 = \frac{1}{2} \int\limits_0^1 dt\; (1-t)^{\frac{d-3}{2}-1} = \frac{1}{d-3}, 
\end{eqnarray}
provided that Re$((d-3)/{2})>0$. {From ~(\ref{regionIII}), one gets}
\begin{eqnarray}
\label{2point-rij0-a}
\hspace{-0.6cm}\dfrac{(d-3)\; J_2^{(d)} }{ \Gamma\Big(2-\frac{d}{2}\Big)}= 
\Bigg\{\dfrac{\partial_j \lambda_{ij}}{4p_{ij}^2}\; \Big(m_i^2\Big)^{\frac{d}{2}-2 }
 + (i \leftrightarrow j)\Bigg\}.
\end{eqnarray}
\subsection{Three-point functions}

\noindent
In a similar manner we can write a complete formula for $J_3^{(d)}$ as
\begin{eqnarray}
\label{3point-solution}
J_3^{(d)} &=& 
\sqrt{8}\; \pi\Gamma\left(2-\frac{d}{2} \right) 
\frac{
\sqrt{-g_{ijk}}}{\lambda_{ijk}}\; (r_{ijk})^{\frac{d-2}{2}} \Theta(\tau_{ijk}) 
\nonumber\\
&&+C_{ijk}^0 +C_{ikj}^0+C_{kji}^0.
\end{eqnarray}
The first term in (\ref{3point-solution}) derives from the boundary condition solving
(\ref{npoint}) also defining the function $\Theta(\tau_{ijk})$, see~\cite{higheps}. 
Depending on the problem it can be chosen in different ways. The functions $C^0_{ijk}$
read
\begin{eqnarray}
\label{c0} 
&&\hspace{-1.5cm} \dfrac{C^0_{ijk}}{\Gamma\left(2-\frac{d}{2} \right) } 
= -\dfrac{ \sqrt{\pi} \; \Gamma\left(\frac{d}{2}-1 \right)} {\Gamma\Big(\frac{d-1}{2}\Big)} 
   \Bigg( \frac{~~r_{ij}^{\frac{d}{2}-1}}{2\lambda_{ij} } \Bigg)
   \Bigg(\dfrac{\partial_k \lambda_{ijk}}{2\lambda_{ijk} }\Bigg) 
                         \nonumber\\
&& \hspace{-1.2cm} \times  \Bigg\{ \frac{\partial_i  \lambda_{ij}} {\sqrt{1-\frac{m_j^2}{r_{ij}}}}
           +\frac{\partial_j \lambda_{ij}} {\sqrt{1-\frac{m_i^2}{r_{ij}}}}  \Bigg\}
            \Fh21 \; \Fz{1,\frac{d-2}{2} }{\frac{d-1}{2}} {\frac{r_{ij}}{r_{ijk}} }              \nonumber\\
&&\hspace{0.35cm}+\frac{\Gamma\left(\frac{d}{2}-1 \right)}{\Gamma\Big(\frac{d}{2}\Big)}
                   \Bigg(\dfrac{\partial_k \lambda_{ijk}}{2\lambda_{ijk} }\Bigg)           \\
&&\hspace{-1.2cm} \times \Bigg\{ \frac{\partial_j \lambda_{ij}}{\sqrt{1-\frac{m_i^2}{r_{ij}}}} 
           \dfrac{~~~\Big(m_i^2 \Big)^{\frac{d}{2}-1} }{2\lambda_{ij} }\; F_1\Bigg(\frac{d-2}{2}; 1; 
           \frac{1}{2}; \frac{d}{2}; \dfrac{m_i^2}{r_{ijk}}, \frac{m_i^2}{r_{ij} } \Bigg)         \nonumber \\
&&\hspace{0.35cm} + (i \leftrightarrow j) \Bigg\}. \nonumber
\end{eqnarray}
{Analogously for}
$C_{ikj}^0$ and $C_{kji}^0$.
{Eq.~(\ref{c0}) reproduces Eq.~(74)  in ~\cite{Fleischer:2003rm}.} 
{Eq.~(\ref{c0}) is} 
valid provided that 
$\Big|\frac{m_{i}^2}{r_{ij}} \Big|<1$, $\Big|\frac{r_{ij}}{r_{ijk}} \Big|<1$
and  Re$\Big(\frac{d-2}{2} \Big)>0$. The latter condition is always met when {$d>2$}.
The kinematic variables $r_{ijk}, r_{ij}$ and $m_{i}$, etc., usually do not satisfy
the former conditions. 
Therefore, if the absolute value of the arguments of {$\Fh21$} 
and {the Appell functions $F_1$} in  (\ref{3point-solution}) {are larger than one,}
one has to perform 
 analytic {continuations}~\cite{Slater,olsson}.

The Appell function in ~(\ref{c0}) obeys a simple integral representation~\cite{Slater}
\begin{eqnarray}
\label{f1-int}
\hspace{-0.8cm}F_1\Bigg(\frac{d-2}{2}; 1; \frac{1}{2}; \frac{d}{2}; x,y \Bigg)
= \tfrac{d-2}{2}\int\limits_0^1 \dfrac{du \; u^{\frac{d-4}{2}}}{(1-xu)\sqrt{1-yu}},
\end{eqnarray}
provided that $|x|, \; |y|<1$ and Re$(\frac{d}{2}-1)>0$.

All {the} special cases for $J_3^{(d)}$ {will be} listed and calculated in detail 
in Ref~\cite{higheps}. In the present note we consider the massless example.
We relabel the external momenta as $p_{ij} = p_i$, $p_{jk} = p_j$ and $p_{ki} = p_k$. 
One then confirms that $g_{ijk} =  2\; \lambda(p_i^2, p_j^2, p_k^2)$, with
$\lambda$ being the K\"allen function, 
$\lambda_{ijk} = -2 p_i^2 p_j^2 p_k^2$.  By taking the derivatives of $\lambda_{ijk}$ 
with respect to $m_i^2$ in ~(\ref{yijk}) and setting 
$m_i^2 \rightarrow 0$, we obtain
\begin{eqnarray}
\partial_i \lambda_{ijk} &=& 2 p_j^2 (p_i^2+p_k^2 -p_j^2), \nonumber\\
\partial_j \lambda_{ijk} &=& 2 p_k^2 (p_i^2+p_j^2 -p_k^2), \\
\partial_k \lambda_{ijk} &=& 2 p_i^2 (p_j^2+p_k^2 -p_i^2). \nonumber
\end{eqnarray}

The analytic solution for $J_3^{(d)}$ is given by (\ref{3point-solution}) by replacing
\begin{eqnarray}
\label{c0-massless}
\hspace{-0.5cm} \dfrac{C_{ijk}^0}{ \Gamma\left(2- \frac{d}{2}\right)} 
&=&  \dfrac{\sqrt{\pi} \; \Gamma\left(\frac{d}{2}-1\right)}{\Gamma\left(\frac{d-1}{2} \right)} 
     \left(-\dfrac{p_i^2}{4}\right)^{\frac{d}{2}-2} \;                                     \\
&&\hspace{-1.5cm}\times \left(\dfrac{1}{p_j^2 + p_k^2- p_i^2} \right) 
   \Fh21 \;\Fz{1,\frac{1}{2} }{\frac{d-1}{2}}{\dfrac{\lambda(p_i^2, p_j^2, p_k^2)}{(p_j^2 + p_k^2- p_i^2)^2 } }.  
\nonumber
\end{eqnarray}
For $p_i^2>0$  {one applies $p_i^2+i\epsilon$ in (\ref{c0-massless}) and reproduces} 
the results given in ~\cite{Davydychev:1999mq},
using another method.
\subsection{Four-point functions} 

\noindent
By applying the same procedure, one can write
a compact formula for $J^{(d)}_4$ as follows
\begin{eqnarray}
\label{d0-formular}
J_4^{(d)} &=&8 {\pi}^{\frac32}~\frac{\Gamma\left(\frac{d}{2}\right) \Gamma\left(2 - \frac{d}{2}\right)} {(d-2) \Gamma\left(\frac{d-3}{2}\right)}
\frac{\sqrt{-g_{ijkl}}}{\lambda_{ijkl}}~ r_{ijkl}^{\frac{d-3}{2}} \; \Theta (\tau_{ijkl}) \nonumber\\
&&+ D^0_{ijkl}+D^0_{lijk}+D^0_{klij}+D^0_{jkli},
\end{eqnarray}
with 
\begin{eqnarray}
\label{dijkl0}
&&\hspace{-1.6cm} \dfrac{D^0_{ijkl} }{ \Gamma\left(2-\frac{d}{2}\right)} 
=  -\sqrt{8} \pi \frac{\sqrt{-g_{ijk}}}{\lambda_{ijk}}\; \Big(r_{ijk} \Big)^{\frac{d-2}{2}}                         
\nonumber\\
&&\hspace{0.2cm}\times \Bigg(\dfrac{\partial_l \lambda_{ijkl}}{2\lambda_{ijkl}}\Bigg)
\Fh21 \; \Fz{\frac{d-3}{2} ,1}{\frac{d}{2} -1  }{\frac{r_{ijk}}{r_{ijkl}} }\; \Theta (\tau_{ijk})  \nonumber\\
&&\hspace{-1.6cm} + \Bigg\{ \frac{\sqrt{\pi}\;\Gamma\left(\frac{d}{2}-1\right) }
 {\Gamma\left(\frac{d-1}{2}\right) } \left(\dfrac{\partial_l \lambda_{ijkl}}{2\lambda_{ijkl}}\right)
 \Bigg(\dfrac{\partial_k \lambda_{ijk}}{2\lambda_{ijk} }\Bigg)                                \nonumber\\
&&\hspace{0.2cm}  \times \Bigg( \frac{r_{ij}^{\frac{d}{2}-1}}{2\lambda_{ij} } \Bigg)
\Bigg( \frac{\partial_i  \lambda_{ij}}{\sqrt{1-\frac{m_j^2}{r_{ij}}}} \Bigg)
\Bigg( \frac{1}{\sqrt{1-\frac{r_{ij} }{r_{ijk}}}} \Bigg)                                    \\
&&\hspace{0.8cm}  \times F_1\left(\frac{d-3}{2}; 1, \frac{1}{2}; \frac{d-1}{2}; 
 \frac{r_{ij}}{r_{ijkl}},\frac{r_{ij}}{r_{ijk}} \right)                                              \nonumber \\ 
&&\hspace{-1.6cm} -\dfrac{ \Gamma\left(\frac{d}{2}-1\right)}{\Gamma\Big(\frac{d}{2}\Big)}
\Bigg(\dfrac{\partial_l \lambda_{ijkl}}{2\lambda_{ijkl}}\Bigg)
\Bigg(\dfrac{\partial_k \lambda_{ijk}}{2\lambda_{ijk} } \Bigg) 
\Bigg( \frac{\partial_j \lambda_{ij}} {2\lambda_{ij}}   \Bigg)                               \nonumber\\
&&\hspace{0.2cm} \times \Bigg(\dfrac{r_{ijk}}{r_{ijk}-m_i^2} \Bigg)
\Bigg(\dfrac{r_{ij}}{r_{ij}-m_i^2} \Bigg) \Big(m_i^2\Big)^{\frac{d}{2}-1} \times                     \nonumber\\
&& \hspace{-1.6cm}\; F_S\Bigg(\frac{d-3}{2}, 1, 1; 1,1,\frac{1}{2}; 
 \frac{d}{2},\frac{d}{2},\frac{d}{2};\frac{m_i^2}{r_{ijkl}},\dfrac{m_i^2}{m_i^2-r_{ijk}}, 
\frac{m_i^2}{m_i^2-r_{ij} } \Bigg)                                                                   \nonumber\\
&&\hspace{-1.5cm}\quad + (i \leftrightarrow j)\;\;\; \Bigg\}                                         \nonumber \\
&&\hspace{-1.5cm} + \Big\{(i,j,k) \leftrightarrow (k,i,j)\Big\} 
+ \quad \Big \{(i,j,k) \leftrightarrow (j,k,i)\Big\}.                                                \nonumber
\end{eqnarray}
{Eq.~(\ref{dijkl0}) reproduces Eq.~(99)  in ~\cite{Fleischer:2003rm}.} 
The terms $ D^0_{lijk},\; D^0_{klij},\;D^0_{jkli}$ are obtained  
from $D^0_{ijkl}$ by circular permutation of the {indices} $i,j,k,l$, 
and the first first term in (\ref{d0-formular}) results form the boundary condition,
cf.~\cite{higheps} for details.
The representation for $J_4^{(d)}$ in (\ref{d0-formular}) with $D_{ijkl}^{0}$, and 
equivalently for $ D^0_{lijk},\; D^0_{klij},\;D^0_{jkli}$ in ~(\ref{dijkl0}), 
is valid under the conditions that Re$\Big(\frac{d-3}{2}\Big)>0$ and {that} the absolute values
of arguments of the hypergeometric functions {are smaller than one}. If the absolute value of 
the arguments {are larger than 
one,} 
one has to perform  analytic continuations of the hypergeometric  and Appell 
$F_1$ functions, cf.~\cite{Slater,olsson}. {Further, the Saran function $F_S$ may be expressed}
by a Mellin-Barnes representation in this case.

\noindent 
The integral representation of $F_S$ reads \cite{Saran55}
\begin{eqnarray}
\label{lauricella-int}
&&\hspace{-1.2cm}F_S(\alpha_1,\alpha_2,\alpha_2,
       \beta_1, \beta_2, \beta_3,
       \gamma_1,\gamma_1,\gamma_1; x,y,z) =\\
&&\hspace{-1cm}=\frac{\Gamma (\gamma_1)}
   {\Gamma (\alpha_1 ) \Gamma (\gamma_1-\alpha_1)}
   \int_0^1  dt\; \dfrac{t^{\gamma_1-\alpha_1-1}(1-t)^{\alpha_1-1}}
   {(1-x+tx)^{\beta_1}}  \nonumber\\
&& \hspace{1.7cm}\times 
   F_1\left(\alpha_2,\beta_2,\beta_3;\gamma_1-\alpha_1; ty,tz\right), \nonumber
\end{eqnarray}
provided that $|x|,\;|y|,\; |z|<1$ and Re$(\gamma_1-\alpha_1-\alpha_2)>0$.

All the special cases are treated in  Ref~\cite{higheps} in detail. 
Again we consider the massless case as an example. We perform the analytic continuation of the result 
for $J_4^{(d)}$ in (\ref{d0-formular},~\ref{dijkl0}) in this case. 
Furthermore, taking $m_i^2 \rightarrow 0$, 
one notices that 
\begin{eqnarray}
&&\hspace{-1.2cm}F_S\left( \frac{d-3}{2},1,1; 1,1,\frac12;
\frac{d}{2},\frac{d}{2},\frac{d}{2};0,0,0 \right)= 1,
\end{eqnarray}
provided that Re$\Big(\frac{d-3}{2} \Big)>0$.  On the other hand, 
$ (m_i^2)^{\frac{d}{2}-1} \rightarrow 0 $ whenever $d>2$. 
Therefore the terms related to $F_S$ in ~(\ref{dijkl0}) are vanishing {in the massless case}.
As a result, the term $D^0_{ijkl}$ can be written as
\begin{eqnarray}
\label{dijkl0-mi0}
&&\hspace{-1.6cm} \dfrac{D^0_{ijkl} }{ \Gamma\left(2-\frac{d}{2}\right)} 
=  -\sqrt{8} \pi \frac{\sqrt{-g_{ijk}}}{\lambda_{ijk}}\; \Big(r_{ijk} \Big)^{\frac{d-2}{2}}                         
\nonumber\\
&&\hspace{0.2cm}\times \Bigg(\dfrac{\partial_l \lambda_{ijkl}}{2\lambda_{ijkl}}\Bigg)
\Fh21 \; \Fz{\frac{d-3}{2} ,1}{\frac{d}{2} -1  }{\frac{r_{ijk}}{r_{ijkl}} }\; \Theta (\tau_{ijk})  
\nonumber\\
&&
 +\frac{\sqrt{\pi} \;\Gamma\left(\frac{d}{2}-1\right) }
      {2\; \Gamma\left(\frac{d-1}{2}\right) } 
      \left(\dfrac{\partial_l \lambda_{ijkl}}{2\lambda_{ijkl}}\right)
      \Bigg(\dfrac{\partial_k \lambda_{ijk}}{2\lambda_{ijk} }\Bigg)  
      \nonumber \\
&&\hspace{-1.3cm}\times   
    \Bigg( \frac{r_{ij}^{\frac{d}{2}-2} }{\sqrt{1-\frac{r_{ij} }{r_{ijk}}}} \Bigg)
    \; F_1\left(\frac{d-3}{2}; 1, \frac{1}{2}; \frac{d-1}{2}; 
    \frac{r_{ij}}{r_{ijkl}},\frac{r_{ij}}{r_{ijk}}    \right) \nonumber\\
&& \hspace{-1.3cm}+ \Big\{ (i,j,k) \leftrightarrow (j,k,i) \Big\} 
    +\Big\{ (i,j,k) \leftrightarrow (k,i,j) \Big\}. 
\end{eqnarray}
The $\varepsilon$-expansion of all the above expressions can be performed using the packages
{\tt Sigma, EvaluateMultiSums} and {\tt Harmonic Sums} \cite{PACKAGES}. Compact expressions for the 
{numerics of} 5-point and higher{-point} functions are given in ~\cite{Fleischer:2010sq,Fleischer:2011hc}.

\vspace*{-3mm}\noindent
\section{Numerical checks}

\noindent
The analytic results have been implemented as 
{\tt Mathematica v$8.0$} package {\tt ONELOOP234.m}.
In Tables~1--5 we compare {\tt ONELOOP234.m} 
with {\tt AMBRE/MB v$1.2$} and {\tt LoopTools/FF v.2.10}. The results show a very good agreement in all cases.

\vspace*{-3mm} \noindent
\section{Conclusions}

\noindent
A systematic study of the scalar one-loop 
two-, three- and four-point integrals in arbitrary space-time dimensions is presented. {For} considering all cases of 
mass- and 
external invariant assignments, {and 
for the systematic derivation of proper $\varepsilon$ expansions we refer to} ~\cite{higheps}. We 
performed numerical {sample} checks {with} {\tt LoopTools/FF}
and {\tt AMBRE/MB}, finding {perfect} agreement. 
Packages both in {\tt Mathematica} and {\tt Fortran}
providing the corresponding expressions will be {made} available in the near future.

\vspace{1mm}
\noindent
{\bf Acknowledgment.}~
This work was supported in part by the European
Commission through contract PITN-GA-2012-316704 ({HIGGSTOOLS}).

\clearpage

\vspace*{-10mm} \noindent
\begin {table}[H] 
\caption{Comparison of {the $\varepsilon^0$ term of}  $J_2^{(4-2 \varepsilon)}$ with {\tt LoopTools/FF.}}
\begin{center}
\scriptsize  
\begin{tabular}{|l|l|}\hline
$(p^2,m_1^2,m_2^2)$  & This work/{\tt LoopTools/FF} \\ \hline \hline
$(1000, 25, 36)$       & $-4.6291944000901360  - 2.9439271197312254\; i $    \\ \cline{2-2}
                       & $-4.6291944000901353  - 2.9439271197312253\; i $    \\ \hline
$(-1000,25,36)$      & $ -5.1755714223812398 + 0.0\;i$                     \\ \cline{2-2}
                       & $ -5.1755714223812399 + 0.0\;i$                     \\ \hline \hline
\end{tabular}
\end{center}
\end{table}


\vspace*{-9mm} \noindent
\begin {table}[H] 
\caption{Comparison of {the $\varepsilon^0$ term of} $J_3^{(4-2 \varepsilon)}$ with {\tt LoopTools/FF}, 
with $m_1^2=16$, $m_2^2=25$, $m_3^2=36$.}
\begin{center}
\scriptsize  
\begin{tabular}{|l|l|}\hline
$(p_1^2, p_2^2, p_3^2)\times 10^2 $  & This work/{\tt LoopTools/FF} ($\times 10^{-2}$) \\ \hline \hline
$(1, 5, 2)$      & $ 1.62458585289425548 - 0.81633722940756023\; i$           \\  \cline{2-2}
                 & $ 1.62458585289425653 - 0.81633722940756131\; i$           \\ \hline
$(-1, -5, -2)$   & $ -0.622465213238832772 + 0.0 \; i$                        \\ \cline{2-2}
                 & $ -0.622465213238832665 + 0.0 \; i$                        \\ \hline \hline
\end{tabular} 
\end{center}
\end{table}

\vspace*{-14mm} \noindent
\begin {table}[h] 
\caption{Comparison  {of the $\varepsilon^0$ term} of $J_4^{(4-2 \varepsilon)}$ with
{\tt LoopTools/FF} in the massless case.}
\begin{center}
\footnotesize
\begin{tabular}{|l|l|}\hline
$(p_1^2, p_2^2, p_3^2, p_4^2, {s, t})$  & This work/{\tt LoopTools/FF}       \\ \hline \hline
$(1, 5,1,7, 15, 1)$      & $ 0.22052449908760818 + 0.0\; i$  \\ \cline{2-2}
                         & $ 0.22052449908760813 + 0.0\; i$  \\ \hline
${(-1, -5, -1, -7, -25, -1)}$   & $ {0.18243214608579919 + 0.0\; i}$  \\ \cline{2-2}
                         & $ {0.18243214608579872 + 0.0\; i}$  \\ \hline \hline
\end{tabular}
\end{center}
\end{table}

\vspace*{-11mm} \noindent
\begin {table}[H] 
\caption{Comparison of  $J_2^{(12-2 \varepsilon)}(-100; 100, 400)$ $\times 10^{{8}}$
with {\tt  AMBRE/MB}. The Monte Carlo error of {\tt AMBRE/MB} is $\mathcal{O}(10^{-1{2}})$.}
\begin{center}
\scriptsize
\begin{tabular}{|l|l|}\hline \hline
This
& $+  3.42724867724867725 \varepsilon^{-1} - 12.8050782926747736   \varepsilon^0$  \\
work
& $+ 29.2335688024947187  \varepsilon^1  - 49.6610273102650704  \varepsilon^2$    \\
& $+ {70.9545648069212449}\varepsilon^3 {- 89.4570353152169962}  \varepsilon^4$    \\ \hline
{\tt AMBRE} 
&$ {+3.427248677248677  \varepsilon^{-1} - 12.805078292674778 \varepsilon^0}$\\
&${+ 29.23356880249472\varepsilon^1 -  49.66102731026506\varepsilon^2}$     \\
&$ {+ 70.95456480692126\varepsilon^3 - 89.457035315217\varepsilon^4}$ \\ \hline \hline
\end{tabular}
\end{center}
\end{table}

\vspace*{-10mm} \noindent
\begin {table}[H] 
\caption{Comparison of $J_3^{(12-2 \varepsilon)}(-100, -500, -200; 16, 25, 36)$ with {\tt AMBRE.} 
The number{s} in bracket{s} show the Monte Carlo error of {\tt AMBRE/MB}.}
\begin{center}
\scriptsize  
\begin{tabular}{|l|l|}\hline \hline
This
&$ +86945.7178571428571\;\varepsilon^{-1} - 252266.637358344924\; \varepsilon^0$     \\ 
work
&$ +499059.587505861549\; \varepsilon^1-741026.396679848443\;\varepsilon^2$          \\
&$ +968401.630700711348\;\varepsilon^3 -1.12777896348311805\cdot 10^6\;\varepsilon^4$\\ \hline
{\tt AMBRE}
& $ +86945.7\; \varepsilon^{-1} -252265.\; \varepsilon^0\;(\pm 7.) 
    +499054.\; \varepsilon\; (\pm 46.)$                                  \\
& $ - 740919.\; \varepsilon^2 \;(\pm  249.) 
    +967850.\; \varepsilon^3\;(\pm 1256.)$                               \\
&$  -1.12699\cdot 10^6 \;\varepsilon^4 (\pm 6398.)  $   
            \\\hline \hline
\end{tabular}
\end{center}
\end{table}




\end{document}